\documentclass[fleqn,twoside]{article}
\usepackage{espcrc2}

\usepackage{graphicx}

\usepackage{latexsym} 
\usepackage{amssymb}  

\newcommand{\AmS}{{\protect\the\textfont2
  A\kern-.1667em\lower.5ex\hbox{M}\kern-.125emS}}

\newcommand{\al}{\alpha}
\newcommand{\be}{\beta}
\newcommand{\ga}{\gamma}
\newcommand{\Ga}{\Gamma}
\newcommand{\de}{\delta}
\newcommand{\De}{\Delta}

\newcommand{\ka}{\kappa}

\newcommand{\La}{\Lambda}

\newcommand{\om}{\omega}

\newcommand{\p}{\partial}
\newcommand{\<}{\langle} 
\renewcommand{\>}{\rangle} 

\newcommand{\txt}{\textstyle}

\newcommand{\dsp}{\displaystyle}

\newcommand\eqn[1]{(\ref{#1})}      

\newcommand{\beq}{\begin{equation}}
\newcommand{\eeq}{\end{equation}}
\newcommand{\ba}{\begin{array}}
\newcommand{\bea}{\begin{eqnarray}}
\newcommand{\ea}{\end{array}}
\newcommand{\eea}{\end{eqnarray}}
\newcommand{\bi}{\begin{itemize}}  
\newcommand{\ei}{\end{itemize}}
\newcommand{\ben}{\begin{enumerate}} 
\newcommand{\een}{\end{enumerate}}
\newcommand{\bc}{\begin{center}}
\newcommand{\ec}{\end{center}}
\newcommand{\bl}{\begin{flushleft}}
\newcommand{\el}{\end{flushleft}}
\newcommand{\br}{\begin{flushright}}
\newcommand{\er}{\end{flushright}}

\newcommand\comment[1]{ \hbox{[{\it Comment suppressed here.}\/]} }
\newcommand\hide[1]{}

\renewcommand{\O}{{\cal O}}
\newcommand{\tr}{\hbox{tr}}

\newcommand{\skipover}[1]{}

\newcommand{\half} {{\txt \frac{1}{2}}}

\newcommand{\quarter}{{\txt\frac{1}{4}}}

\newcommand{\percent}{\symbol{'045}}

\newcommand{\fm}{{\rm fm}} 
 
\newcommand{\keV}{{\rm keV}} 
\newcommand{\MeV}{{\rm MeV}} 
\newcommand{\GeV}{{\rm GeV}} 
\newcommand{\Qt}{{\tilde Q} }
\newcommand{\qbar}{{\bar q}}
\newcommand{\psibar}{{\bar \psi}}
\newcommand{\Msolar}{M_\odot}
\newcommand{\Kelvin}{{\rm K}}

\title{QCD at high density/temperature}

\author{M. Alford\address{Physics and Astronomy Department \\
        Glasgow University \\ Glasgow G12 8QQ \\ U.K.}}
       
\begin{document}

\begin{abstract}
I summarize recent work on QCD at high temperature and density.
\vspace{1pc}
\end{abstract}


\maketitle

\newcommand{\preprintno}{
  \normalsize GUTPA/02/09/02
}

\section{INTRODUCTION}
\label{sec:intro}

QCD now occupies an established position as the 
well-attested theory 
of strong-interaction physics.
Other contributions to these proceedings
(those of S.~Frixione, L.~Lellouch, and
K.~Long) survey the agreement between 
evidence from particle collider experiments and
perturbative and lattice QCD calculations.

However, there is more to QCD than collisions between 
pairs of particles.
QCD should also be able to tell us about the
thermodynamics of matter in the realm of extraordinarily high temperatures
\mbox{($\gtrsim100~\MeV$)}
and densities at which it comes to dominate the behavior.
This regime has real physical importance:
the whole universe was hotter than $100~\MeV$ for the
first crucial microseconds of its history, and in the current epoch
our galaxy has a large population of
neutron stars consisting of matter
squeezed beyond nuclear density by gravitational forces.
However, only in the last few years have these regimes begun to be probed
experimentally in heavy ion collisions 
and astrophysical observations
of neutron stars, and we are still working towards achieving a proper
theoretical understanding of them.

In this paper I will survey the progress that has been 
made in the last few years in qualitative predictions
and quantitative calculations of the properties
of QCD at very high temperature and net baryon density.
I will not discuss theory that is specifically tailored to
heavy ion collisions, but refer the reader to existing review
articles \cite{heavyion}. Other more tightly
focused reviews of QCD at high density  \cite{DenseReviews}
or high temperature \cite{HotReviews} may also prove
useful for details that go beyond the scope of
this summary.

\section{Phases of QCD}

\begin{figure}[t]
\includegraphics[width=0.9\hsize]{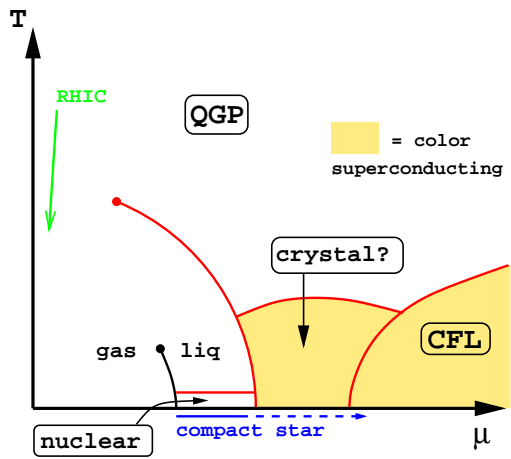}
\caption{A conjectured phase diagram for QCD in the real world.
The CFL region may alternatively extend down to the nuclear phase,
with no intermediate phase. Depending on the strength of instanton 
interactions, the CFL phase may include $K^0$ condensation.}
\label{fig:phase}
\end{figure}

Fig.~\ref{fig:phase} shows a conjectured phase diagram for QCD.
Along the horizontal axis the temperature is zero, and
the density rises from the onset of nuclear matter through the
transition to
quark matter. Compact stars are in this region of the phase diagram,
although it is not known whether their cores are dense enough
to reach the quark matter phase.
Along the vertical axis the temperature rises, taking
us through the crossover from a hadronic gas to the quark gluon plasma.
This is the regime explored by high-energy heavy-ion colliders such
as RHIC (see the contribution by J.~Harris to these proceedings).
At high temperature and moderate density we find the first-order
chiral phase transition ending at the chiral
critical point \cite{BergesRajagopal}, which could perhaps be
detected \cite{SRS-99} in medium-energy heavy ion colliders 
(see Sect.~\ref{sec:hiden-latt}).

For some parts of the phase diagram reliable calculations can
be performed. Standard perturbative
methods can be used in the outer regions of high 
density ($\mu\gtrsim 10^8~\MeV$ \cite{RajagopalShuster})
and temperature ($T\gtrsim 10^{387}~\GeV$ \cite{az}),
and resummation methods
can extend this down to $T\gtrsim 3 T_c$ (Section \ref{sec:pert}).
At zero density and $T\lesssim 4T_c$, lattice methods give good results,
and ways have recently been found to extend lattice methods
to $\mu\lesssim T$. However, this leaves much of the most interesting
phase structure in regions where controlled calculations are not yet
possible. In the following sections, I will describe the progress that
has been made in understanding QCD at high density,
in achieving first-principles calculations at high temperature, 
in lattice calculations
at high temperature and low density, and in
gaining control of field theories out of
thermal equilibrium.

\section{High density QCD}

It is difficult to make accurate
predictions for high density QCD because lattice gauge theory 
is stymied by the complexity of
the fermion determinant (see Sect.~\ref{sec:hiden-latt}). 
However, we can borrow some intuition from
condensed matter physics, which tells us to expect the very
interesting phenomenon of 
color superconductivity 
\cite{oldcolorSC,newcolorSC}
QCD is asymptotically free---the interaction becomes weaker as the
momentum grows---so at sufficiently high density and low temperature,
there is a  Fermi surface of almost free quarks. 
The interactions between
quarks near the Fermi surface are certainly attractive in some channels
(after all, QCD binds quarks together to form baryons)
and it was shown by Bardeen, Cooper, and
Schrieffer (BCS) \cite{BCS} that if there is {\em any} channel in which the
interaction is attractive, then there is a state
of lower free energy than a simple Fermi surface. That state arises
from  a complicated coherent 
superposition of pairs of particles (and holes)---``Cooper pairs''.

Heuristically, the BCS argument is this. 
The Helmholtz free energy is $F= E-\mu N$, where $E$ is
the total energy of the system, $\mu$ is the chemical potential, and
$N$ is the number of fermions. The Fermi surface is defined by a
Fermi energy $E_F=\mu$, at which the free energy is minimized, so
adding or subtracting a single particle costs zero free energy. 
Now switch on a weak attractive interaction.
It costs no free energy to
add a pair of particles (or holes), and the attractive
interaction between them then lowers the free energy of the system.
Many such pairs will therefore
be created in the modes near the Fermi surface, and these pairs,
being bosonic, will form a condensate. The ground state will be a
superposition of states with all numbers of pairs, breaking the
fermion number symmetry. 

Since pairs of quarks cannot be color singlets,
the resulting condensate will break the local color symmetry
$SU(3)_{\rm color}$.  We call this ``color superconductivity''.
Note that the quark pairs play the same role here as the Higgs particle
does in the standard model: the color-superconducting phase
can be thought of as the Higgs phase of QCD.

\subsection{Calculation methods}

To decide whether or not fermions condense in the ground state, one
can explicitly construct a wavefunctional with the appropriate
kind of pairing, and use a many-body variational approach to
find the pairing strength by minimizing the free energy.
Alternatively but equivalently, one can formulate
the problem field-theoretically, in terms of
mean-field Schwinger-Dyson equations
for a quark self energy that includes ``anomalous'' $\<qq\>$
terms as well as the usual $\<\qbar q\>$ terms.
In either case the problem reduces to solving
a self-consistency equation, the gap equation,
to find the self energy. 
If the only solution is zero, there is no condensation in that channel.
If not,
there can be condensation, but it may just be a local minimum of the
free energy.  There may be other solutions to the gap equation, and
the one with the lowest free energy is the true ground state.

There are several possible choices for the interaction to be used
in the Lagrangian from which we derive the
gap equation. At asymptotically high densities
($\mu\gtrsim 10^8~\MeV$ \cite{RajagopalShuster})
QCD is weakly coupled, so one gluon exchange is appropriate. 
Such calculations 
\cite{pert,3flavpert}
demonstrate from first principles that color superconductivity
occurs in QCD.
However, the density regime of physical interest for neutron stars
or heavy ion collisions is 
up to a few times nuclear density ($\mu \lesssim 500~\MeV$), so
one must use a phenomenological interaction that
can be argued to capture the essential physics of QCD.
The usual choice is an NJL model, with a four-fermion interaction
based on known QCD interactions (gluon exchange, or instantons)
whose strength can be normalized to reproduce
known low-density physics such as the chiral condensate, and
then extrapolated to the desired chemical potential.
Such calculations
\cite{oldcolorSC,newcolorSC,CarterDiakonov,ARW3}
find that the color superconducting gap is insensitive to the
form of the interaction, and rises to values around $100~\MeV$.
Various other calculations that include competition
between chiral symmetry breaking and quark pairing
confirm this
\cite{BergesRajagopal,coupledgap}. Very recently,
the first lattice calculation of 
a 3+1 dimensional NJL model with chemical potential
has been performed, and there 
are indications of diquark condensation
\cite{3+1GrossNeveu}.

The favored pairing pattern at high densities, where many 
strange quarks are present, is ``color-flavor locking'' (CFL), which will
be described below. At lower densities
there may be a direct transition from nuclear
matter to CFL quark matter (see Section~\ref{sec:interface}).
Alternatively, at intermediate densities
there may be another phase.
One possibility is the ``2SC'' phase which
has $u$-$d$ pairing only, but this now seems unlikely
to occur in equilibrated quark matter (Section~\ref{sec:2sc}).
Other possibilities include
crystalline pairing (see Section~\ref{sec:loff}), or some
sort of single-flavor pairing \cite{oneflav}.

\subsection{Three flavors: Color-flavor locking (CFL)}
\label{sec:CFL}

In QCD with three flavors of massless quarks 
the Cooper pairs {\em cannot}
be flavor singlets, and both color and flavor symmetries are
necessarily broken \cite{ARW3}.
Both NJL \cite{ARW3,SW-cont,HsuCFL} and gluon-mediated interaction calculations
\cite{3flavpert} agree that the attractive channel 
exhibits a pattern
called color-flavor locking (CFL)
\footnote{Interestingly, such a pattern was at one time
considered as a possibility for 
symmetry breaking at zero density\cite{SS}.
},
\beq
\ba{rcl}
\< q^\al_{i} q^\be_j \> &\sim& \de^\al_i\de^\be_j + \ka\, \de^\al_j\de^\be_i
  \\[2ex]
\multicolumn{3}{c}{
 {[SU(3)_{\rm color}]}
 \times \underbrace{SU(3)_L \times SU(3)_R}_{\dsp\supset [U(1)_Q]}
 \times U(1)_B 
 } \\[3ex]
&\to& \underbrace{SU(3)_{C+L+R}}_{\dsp\supset [U(1)_{\Qt}]} 
  \times \mathbb{Z}_2
\ea
\eeq
(color indices $\al,\be$ and flavor indices 
$i,j$ run from 1 to 3).
The Kronecker deltas connect
color indices with flavor indices, so that the condensate is not
invariant under color rotations, nor under flavor rotations,
but only under simultaneous, equal and opposite, color and flavor
rotations. Since color is only a vector symmetry, this
condensate is only invariant under vector flavor+color rotations, and
breaks chiral symmetry. The effect of including a 
sufficiently large strange quark
mass is to unpair the strange quark \cite{ABR2+1,SW-cont,AR-02}, and at lower
values it may induce a flavor rotation of the condensate
known as ``kaon condensation'' (Ref.~\cite{BedaqueSchaefer}
and Section~\ref{sec:other}).

The features of this pattern of condensation are
\begin{itemize}
\setlength{\itemsep}{-0.7\parsep}
\item The color gauge group is completely broken. All eight gluons
become massive. This ensures that there are no infrared divergences
associated with gluon propagators.
\item
All the quark modes are gapped. The nine quasiquarks 
(three colors times three flavors) fall into an ${\bf 8} \oplus {\bf 1}$
of the unbroken global $SU(3)$, so there are two
gap parameters. The singlet has a larger gap than the octet.
\item 
A rotated electromagnetism (``$\Qt$'')
survives unbroken. It is a combination
of the original photon and one of the gluons.
\item Two global symmetries are broken,
the chiral symmetry and baryon number, so there are two 
gauge-invariant order parameters
that distinguish the CFL phase from the QGP,
and corresponding Goldstone bosons which are long-wavelength
disturbances of the order parameter. 
If a quark mass were introduced then it would explicitly break
the chiral symmetry and give a mass
to the chiral Goldstone octet, but the CFL phase would still be 
a superfluid, distinguished by its baryon number breaking.
\item
The symmetries of the
3-flavor CFL phase are the same as those one might expect for 3-flavor
hypernuclear matter \cite{SW-cont}, so it is possible that there is
no phase transition between them.
\end{itemize}

\subsection{Absence of 2SC}
\label{sec:2sc}

Until recently it was thought that if there were not enough strange
quarks present in quark matter, as might happen at intermediate density,
the favored pairing pattern would
be 2SC, in which the $u$ and $d$ quarks of two colors pair.
leaving the $s$ quarks and third color quarks unpaired.
But in a real neutron star 
we must require electromagnetic and color neutrality
\cite{BaymIida,AR-02}
(ignoring charge-separated phases, see 
Section~\ref{sec:interface}).
It turns out that this penalizes the 2SC phase relative to the
CFL phase \cite{AR-02}.
The reason is that the CFL phase has already paid most of
the cost of neutrality, since it brings the $u$, $d$, and $s$
Fermi surfaces close together
\cite{AR-02,CFLneutral,Steiner:2002gx}, and the 2SC phase
only pairs  4 of the 9 quark colors and flavors, so it
has much less pair binding energy than the CFL phase.

The arguments made in Ref.~\cite{AR-02} are model-independent, based
on simplified assumptions about the dependence of the constituent
strange quark mass $M_s$ on $\mu$ and expanding the free energy in
powers of $M_s/\mu$.  The NJL calculation of
Ref.~\cite{Steiner:2002gx} handles $M_s\sim\mu$ and
includes the coupling between the chiral
condensate and quark condensate gap equations.
The net result is that once neutrality is imposed, there is
no (or very little \cite{Steiner:2002gx}) 
density range in which 2SC is the phase with
the lowest free energy. 

\subsection{Signatures in compact stars}

The only place in the universe where we expect sufficiently
high densities and low temperatures is compact stars,
also known as ``neutron stars'', since it is often
assumed that they are made primarily of neutrons.
A compact star is produced in a supernova.
Typical compact stars have masses
close to $1.4 \Msolar$, and are believed to have radii of order 10 km.
although uncertainty about the equation
of state leaves us unsure of the radius and the maximum density
attained in the core.
During the supernova, the core collapses, and its gravitational energy
heats it to temperatures of order $10^{11}~\Kelvin$ (tens of \MeV),
but it cools rapidly by neutrino emission.
Within a few minutes its internal temperature $T$ drops to
$10^9$~K ($100~\keV$).
Neutrino cooling continues to dominate for the first
million years of the life of the star.
(For detailed treatments and and reviews see Ref.~\cite{nstar}).

Color superconductivity affects the equation of state
at order $(\De/\mu)^2$. It also gives mass to
excitations around the ground state: it opens
up a gap at the quark Fermi surface, and makes the gluons
massive. One would therefore expect its main physical manifestations
to relate to transport properties, such as mean free paths,
conductivities and viscosities.
In the following sections we will survey some of these
manifestations.

\subsection{Mass-radius relationship}
Although the effects of color superconductivity on the quark matter
equation of state are subdominant, they may have a large effect
on the mass-radius relationship. The reason for this is that
the pressure of quark matter relative to the hadronic vacuum
contains a constant (the ``bag constant'' $B$) that represents
the cost of dismantling the chirally broken and confining
hadronic vacuum,
\beq
p \sim \mu^4 + \De^2\mu^2 - B~.
\eeq
If the bag constant is large enough so that nuclear matter is favored
(or almost favored) over quark matter at $\mu\sim 320~\MeV$, then the
bag constant and $\mu^4$ terms almost cancel, and the superconducting
gap $\De$ may have a large effect on the equation of state and hence
on the mass-radius relationship of a compact star \cite{AlfordReddy}.

\subsection{Interfaces and mixed phases}
\label{sec:interface}

If color superconducting quark matter exists in compact stars,
we expect it to occupy a core region, surrounded by
a mantle of nuclear matter. There may be a sharp interface
between the two, but it is well known that a mixed phase region
is also possible \cite{Glendenning:1992vb}, given that there are multiple
chemical potentials in nuclear matter: for baryon number, 
electric charge, and also color.
A mixed phase region would have distinctive characteristics.
Its transport properties are very different from those
of the uniform CFL state: neutrino mean free paths,
which are hundreds of meters in the CFL phase
at temperatures of a few \MeV
\cite{Reddy:2002xc},
are short in the mixed phase due to coherent 
scattering~\cite{Reddy:2000ad}. 


The nature of the boundary between nuclear and CFL quark matter
has been studied \cite{ARRW}, and it was found that a mixed
phase only occurs if the surface tension of the interface
is less than about $40~\MeV/\fm^2 = 0.2\times (200~\MeV)^3$, 
a fairly small value
compared to the relevant scales $\La_{\rm QCD}\approx 200~\MeV$,
$\mu\sim 400~\MeV$.
This means that there may well be a sharp interface rather than a mixed
phase, with a baryon and energy density discontinuity of
about a factor of two across it.
Such an interface may modify the
mass vs.~radius relationship for compact stars with quark
matter cores.  It may also have qualitative effects on
the gravitational wave profile emitted during the in-spiral
and merger of two compact stars of this type. Finally, it
will affect the $r$-mode spectrum (see Section~\ref{sec:rmode}), 
and the damping forces to which $r$-modes are subject.

\subsection{Glitches and the crystalline color superconductor}
\label{sec:loff}

There is a crystalline form of quark pairing
(the ``LOFF'' \cite{LOFF} phase) that occurs 
 when two different
types of quark have sufficiently
different Fermi momenta (because their masses or
chemical potentials are different) that BCS pairing cannot occur
\cite{firstLOFF}\footnote{
At large number of colors there is the possibility of
a separate ``chiral crystal'' phase \cite{RappCrystal},
which is not a superfluid, so it will not lead to glitches.
}. 

Such situations are likely to be generic in nature, where, because of
the strange quark mass, combined with requirements of weak equilibrium
and charge neutrality, all three flavors of quark in general have
different chemical potentials. 

The phenomenology of the crystalline phase has not yet been worked out,
but recent calculations using Landau-Ginzburg effective theory 
indicate that the favored phase may be a face-centered cubic crystal
 \cite{Bowers:2002xr}, with a reasonably large binding energy.
This raises the interesting possibility of glitches in quark matter stars.

Glitches are sudden
jumps in rotation frequency $\Omega$ of a pulsar.
They are thought to arise from pinning of superfluid vortices
in the nuclear matter core to the lattice of nuclei
that forms the crust 
\cite{GlitchModels}.
As the pulsar spins down, the vortices move outwards, and eventually
overcome the pinning forces, and rearrange themselves.
Until recently it was thought that quark matter stars would not glitch,
because quark matter was thought to have no crust or other rigid
structure that could pin the vortices.
Now, however, we can envisage an 
alternative scenario for glitches
(supported only be dimensional estimates at this point)
where a superfluid CFL core
is surrounded by a crystalline layer of quark matter in a LOFF phase.

\subsection{Cooling by neutrino emission}

For its first million or so years,
a compact star cools by neutrino emission. There are many uncertainties in
determining the temperature of a compact star (from its X-ray spectrum)
and its age (inferred from the spindown rate)
but observers are now able to map out cooling curves
\cite{Page,Yakovlev:2000jp,Tsuruta-02}.
The cooling rate is
determined by the heat capacity and emissivity, both
of which are sensitive to the spectrum of low-energy excitations,
and hence to color superconductivity 
\cite{Page,Blaschke,Shovkovy-02,Reddy:2002xc}.

In the CFL phase, all quarks and gluons have gaps $\Delta\gg T$,
electrons are absent \cite{CFLneutral}, and the pseudo-Goldstone
bosons have masses of order tens of \MeV\ \cite{effth}, so the
transport properties are dominated by the only true Goldstone
excitation, the superfluid mode arising from the breaking of the exact
baryon number symmetry.  This means that CFL quark matter has a much
smaller neutrino emissivity and heat capacity than nuclear matter, and
hence the cooling of a compact star is likely to be dominated by the
nuclear mantle rather than the CFL core 
\cite{Page,Shovkovy-02,Jaikumar:2002vg}.  
A CFL core is therefore not expected to be detectable by cooling
measurements, although it has been suggested that
it might slow the first $\approx 100$ years of cooling \cite{Blaschke}.

Other phases such as 2SC or LOFF give large gaps to only
some of the quarks, leaving the others  with no gaps
(or very small gaps in weakly attractive channels)
so they may allow fast cooling, which is severely constrained
by the latest space-based X-ray observations \cite{Tsuruta-02}.
The cooling would proceed quickly, then
slow down suddenly when the temperature fell below
the smallest of the small weak-channel gaps.
If no sign of such behavior
is seen as our observations of compact star temperatures improve
then we will have to conclude that either these phases do not
occur, or their gaps are much larger than we expected.

\subsection{The neutrino pulse at birth}

In the  first minute of a 
supernova, the inner regions cool rapidly, radiating off
much of their energy  as neutrinos,
whose detailed spectrum as a function of time
is determined by the neutrino diffusion properties
of the protoneutron star, and can be detected
in terrestrial experiments such as SuperKamiokande.

Initial calculations by Carter and Reddy~\cite{CarterReddy} found
a striking signature: the core released few neutrinos when it was
close to the critical temperature for color superconductivity
$T\approx T_c$, then a late blast of neutrinos
as their mean free path grew larger than the size of the star
at $T\ll T_c$.
This picture is complicated by  scattering of neutrinos
by Goldstone bosons \cite{Reddy:2002xc}.
Also, the temporal variation of neutrino emission from the core 
over the first few seconds is liable to be blurred
by interactions with the surrounding material.
Detailed numerical simulations of core collapse, including
the effects of quark pairing, are now needed.

\subsection{r-mode instability}
\label{sec:rmode}

An ``$r$-mode'' (short for ``rotational mode'') \cite{firstRmode} is a
bulk flow in a rotating star that radiates away energy and angular momentum
in the form of gravitational waves (for a review,
see Ref.~\cite{Friedman:2001as}).
If the rotation frequency 
of the star is above a critical value, the system becomes
unstable to $r$-modes and quickly spins down to the critical value,
 at which point the  $r$-modes are damped out.
The critical frequency depends on the sources of damping,
such as shear and bulk viscosities, and also
friction at the interface between
the $r$-mode region and adjacent region.
One can therefore rule out certain models for compact stars
on the grounds that they have such low damping that 
they could not support the high rotation rates observed in pulsars:
$r$-mode spindown would have slowed them down.

It is generally believed that ordinary nuclear matter is not
ruled out in this way \cite{Bildsten-99}, but
Madsen~\cite{MadsenRmode} has shown that
color superconductivity, by
creating gaps in the quark excitation spectrum,
suppresses the viscosities by factors of
order $\exp(-\De/T)$, encouraging $r$-mode spindown.
Furthermore, since  CFL matter
contains no electrons \cite{CFLneutral}, there is
no electrostatic cushion to support a crust of normal matter,
which would otherwise have been an additional source of friction.
Madsen found that for a compact star made {\em entirely} of quark matter
in the CFL phase, even a gap as small as $\Delta=1$~MeV
is ruled out by observations of millisecond pulsars.

It remains to extend this calculation to the more generic picture of
a quark matter core surrounded by a nuclear mantle. The friction at
the core-mantle interface may be enough
\cite{Bildsten-99,MadsenRmode} to quash the $r$-modes.
Finally, quark matter may contain a region of LOFF
crystal, and the $r$-modes could be damped
at the edges of that region rather than at the crust.

\subsection{Other phenomenology}
\label{sec:other}

Many other physical phenomena associated with color superconductivity
in quark matter have been investigated.
Small chunks of quark matter (``strangelets'') may exist, and
standard calculations must be modified in that context to
account properly for finite volume effects \cite{Amore}.
Strangelets in the  CFL phase will have a different mass-charge
relationship from those made of unpaired quark matter
\cite{Madsen:2001fu}. At high enough densities $U(1)_A$ breaking
is suppressed, leading to domain walls
bounded by strings \cite{U1strings}.

There are many interesting questions about the influence of the strange
quark mass on color superconductivity. When $m_s=0$ the favored phase
is CFL, as discussed in Section~\ref{sec:CFL}. As the strange quark mass
is increased from zero the CFL pairing becomes harder and harder to
maintain, and at some point it ceases to be the ground state
\cite{ABR2+1}. But
before that the system may respond to the finite $m_s$ by rotating
the quark-quark condensate in flavor space, in a direction that
reduces the number of strange quarks. This can be understood as
condensation of kaons \cite{BedaqueSchaefer}, where by ``kaon'' we mean
the pseudo Goldstone bosons that are
associated with the breaking of the chiral flavor symmetry by the
CFL condensate. The resultant CFL-K$^0$ phase would break isospin and
may also include charged kaons
in the first seconds after the supernova, before
neutrinos escape the star \cite{KaplanReddy,Kryjevski:2002ju}.

It is unclear whether kaon condensation will occur in 
dense quark matter
cores of compact stars.
Instanton and electromagnetic interactions tend to disfavor CFL-K$^0$ over
ordinary CFL, and the magnitude of instanton effects
at the densities of interest is very uncertain \cite{Schaefer:2002gf}.
But CFL-K$^0$ has many interesting features. 
It supports
topological or near-topological solitons:
superconducting K-strings
\cite{Kstrings} and domain walls \cite{Kwalls} whose possible
observable manifestations are yet to be investigated.

\section{High temperature QCD}

We now turn from the horizontal axis of the QCD phase diagram
(Fig.~\ref{fig:phase}) to the vertical axis, and study
QCD at high temperature but low (net) baryon number density.
The standard techniques are lattice gauge theory, perturbation theory,
and a hybrid of the two that is generally known as ``dimensional reduction''.

\subsection{Lattice gauge theory} 
Lattice gauge theory is a brute-force evaluation of the
functional integral over gauge and quark field configurations
that is the theoretical definition of QCD.
It is therefore a fully non-perturbative first-principles
calculation of the predictions of QCD.
To be numerically tractable, the integral must be formulated in
Euclidean time,
\beq
\ba{rcl@{\,}l}
Z &=& \int {\cal D}A & {\cal D}\psibar  {\cal D}\psi
  \,\exp(- \psibar M[A] \psi) \\
 &&& \exp( - S_{\rm gauge}[A]) \\[1ex]
 &=& \int {\cal D}A 
  & \det(M[A])  \exp( - S_{\rm gauge}[A])
\ea
\label{lattZ}
\eeq
where the fermionic matrix $M$ is the usual fermion action 
\beq
 M[A]  = \ga^\mu D_\mu + m
\eeq
In general, lattice gauge calculations require large computational resources.
A typical calculation is formulated in a box of size $V\approx (3~\fm)^3$,
with lattice spacing $a\sim 0.1~\fm$, and with up and down quark masses
$m_{u,d} \gtrsim m_s/4$. To obtain physically relevant results it is necessary
to extrapolate $a\to 0$, $V\to\infty$, $m_{u,d}\ll m_s$.
In addition, not all quantities are accessible to lattice calculations:
spectral functions and
transport coefficients are defined in Minkowski space, and can only be obtained
by some sort of analytic continuation from the lattice results
\cite{spectral}.
In spite of these challenges, lattice gauge calculations have been
very successful in obtaining predictions for the equation of state of QCD
at high temperature.

\begin{figure}[htb]
\includegraphics[width=\hsize]{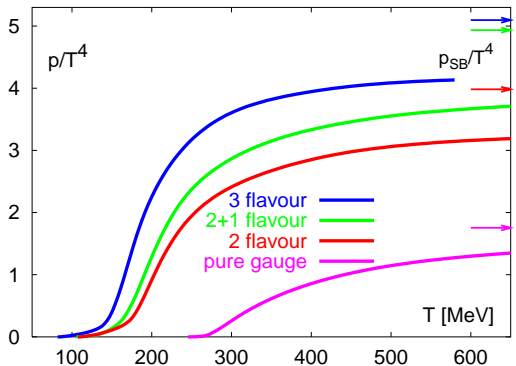}
\caption{The pressure in QCD with various numbers of
dynamical quarks, as a function of temperature. 
Arrows show the corresponding noninteracting quark+gluon gas pressure
$p_{SB}$.
For the (2+1)-flavor calculation the light quark mass was 
$m_{u,d}\approx \quarter m_s$ \cite{Kar00a}.}

\label{fig:latt}
\end{figure}

In a lattice calculation, thermal averages are obtained by setting the
length of the Euclidean time direction to the inverse temperature.
To probe physics near the deconfinement temperature $T_c \sim 200~\MeV
\sim (1~\fm)^{-1}$ therefore requires a lattice spacing
$a \lesssim 0.2~\fm$. To probe much higher temperatures would require
a correspondingly smaller lattice spacing, increasing the numerical
effort. Thus calculations are so far limited to  $T\lesssim 4T_c$.
(This restriction may be eased in the future by the introduction of anisotropic
lattices, which are finer in the $t$-direction.)
State-of-the-art lattice calculations use
improved actions, which include extra terms to compensate for the
$\O(a^2)$ discretization errors, and yield
continuum-like results from relatively coarse 
($a \lesssim 0.2~\fm$) lattices. This is crucial
because the cost of a lattice calculation rises as a very high power of $1/a$.
Using improved staggered \cite{Kar00a} and Wilson \cite{cp-pacs}
actions the equation of state of QCD has been calculated and is
shown in Fig.~\ref{fig:latt}.

The fact that at $T\sim 4T_c$ the pressure
is about 15\percent\ less than the pressure
$p_{SB}$  of free quarks and gluons shows 
that there are still important interactions at that temperature.
The major challenge for other approaches is to
confirm and explain this.

\subsection{Perturbation theory} 
\label{sec:pert}
Perturbation theory is generally applicable when the
coupling is small. One might expect that quarks and gluons
at temperature $T$ would be described by QCD with running coupling
$g(T)$, which becomes small at high temperature, causing perturbation
theory to converge quickly.
In fact, this does not in general happen. 
For example, attempts to calculate
the free energy to order ${\cal O}(g^5T^4)$~\cite{az,zk,bn}
find that there are large corrections with alternating 
signs, making the naive perturbative series useless for 
physically interesting temperatures
(for illustration, the $g^3$ prediction is shown in Fig.~\ref{fig:pert}).
As we will see below, there are resummations of the perturbative series 
that greatly improve its convergence. There
are also now results to leading order in $g$ for transport coefficients,
whose expansion contains powers of $\log(g)$ and therefore converges
even more slowly.

\subsubsection{Equation of state}
\begin{figure}[htb]
\includegraphics[width=\hsize]{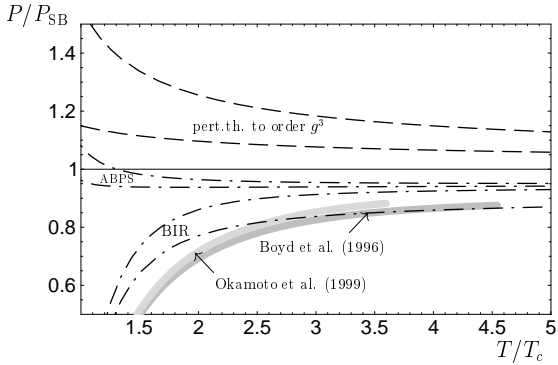}
\caption{
Results \cite{Hwa:future} of calculations of the pressure
of hot gluons, from 
 lattice calculations (Grey bands) and  $\O(g^3)$ naive perturbation theory
(dashed lines). The dot-dashed lines
show the range of predictions from 
the HTL re-orderings of the perturbative series formulated by
Blaizot, Iancu, and Rebhan (BIR), and
Andersen, Braaten, Petitgirard, and Strickland (ABPS).
}
\label{fig:pert}
\end{figure}

There have been various proposals for reorganizing 
the high-temperature perturbative series for the pressure,
in order to improve its convergence. They involve
resumming the hard thermal loops (HTL), typically via the
Cornwall-Touboulis-Jackiw two-particle
irreducible (2PI) effective action \cite{CJT-74},
in which the complete propagator
is used as an infinite set of variational parameters \cite{FM-77,Phi}.
This has been used in 
approximately self-consistent HTL resummation calculations of the entropy by
Blaizot, Iancu and Rebhan (BIR) \cite{BIR}
and of the pressure by Peshier \cite{Peshier-00}.
In a different approach, Andersen, Braaten, Petitgirard, Strickland (ABPS)
used the HTL effective action
with the Debye screening mass as a variational parameter \cite{ABPS-02}.
The results of such calculations are shown in Fig.~\ref{fig:pert}.
The dashed lines show  $\O(g^3)$ naive perturbation theory, 
to illustrate the poor convergence of the perturbative series.
(The  $\O(g^2)$ and $\O(g^4)$ predictions lie below the free gas result.)
The theoretical uncertainty comes from varying the renormalization scale
by a factor of 2 around the first Matsubara frequency $2\pi T$.
The grey bands show the results of lattice calculations \cite{greylatt}.
Finally, the range of predictions of the BIR
 and ABPS hard thermal loop methods are shown.
The BIR prediction has a bigger theoretical uncertainty
because they choose a scale to separate hard and soft contributions,
and varying that scale contributes to the error. BIR are currently working
to reduce this uncertainty.

We conclude that there are (perhaps somewhat ad hoc)
resummations of perturbation theory that indicate that
the pressure should slowly approach the free QGP value $P_{SB}$ from below.
This is in agreement with, but not yet as accurate as, 
the lattice calculations. At this point it appears difficult to
push these calculations to higher order and thereby judge their
validity.

\subsubsection{Transport properties}
The photon emission
rate from a quark-gluon plasma has been perturbatively 
estimated by various
groups \cite{Kapusta,Baier,Gelis1}
to leading order in the electromagnetic and strong
coupling constants. However, there are complications due to 
leading logs that arise from multiple scattering of 
co-linear bremsstrahlung (Landau-Pomeranchuk-Migdal (LPM) effect).
Only recently have these been fully taken into account
by Arnold, Moore, and Yaffe \cite{Arnold:2001ms}. They found that
the soft ($k\lesssim 2T$) photon emission rate behaves as
$d \Ga/ dk \propto k^{-1/2}$, which is less infrared singular
than the result  $d \Ga/ dk \propto k^{-1}$
obtained in earlier calculations that neglected the LPM effect
\cite{Gelis1,softgamma}.
Similar methods are applicable to the
dilepton rate, which is an important observable in
heavy-ion collisions, 
and also to transport properties such as the
shear viscosity, electrical conductivity, and
flavor diffusion constant, which are in principle important
input parameters to phenomenological models of the early
stages of a heavy-ion collision. We can therefore look forward
perturbative calculations of these quantities to leading order in $g$
in the near future.

\subsection{Dimensional reduction} 

The intuition behind dimensional reduction is related to
the poor behavior of the perturbative expansion in QCD at high temperature,
which can be understood by considering the $T\to \infty$
limit of the pure gauge theory. 
At non-zero $T$ the Euclidean time size is finite,
so the glue fields can be Fourier-analyzed into Matsubara modes, with
frequencies $\om = 0,\ 2\pi T,\  4\pi T,$ etc.
As $T\to \infty$, the Euclidean time size goes
to zero, and the $\om=0$ Matsubara ``static mode'' dominates.
The theory reduces to a three-dimensional theory of gluons 
(the spatial components $A_i$)
and a scalar that transforms as a color octet ($A_0$) 
 \cite{bn,dr,rold,hl,generic,ad}
\beq
{\cal L} = 
\half \tr F_{ij}^2
+ \tr [D_i,A_0]^2 + 
m_D^2\tr A_0^2 +\lambda_A(\tr A_0^2)^2
\label{L3D}
\eeq
This effective theory is confining, which indicates that the lowest
Matsubara  mode contains the non-perturbative physics 
\cite{linde,gpy}.
It therefore makes sense to perturbatively integrate out the
higher Matsubara modes, and obtain a 3D effective action for the
static mode. This can then be treated by lattice methods.

The final result for the pressure takes the form of an integral over
a coupling constant in the 3D theory, with a parameter $e_0$
associated with the constant of integration
which can be determined by an arduous
4-loop calculation in the 4D theory.

\begin{figure}[htb]
\includegraphics[width=\hsize]{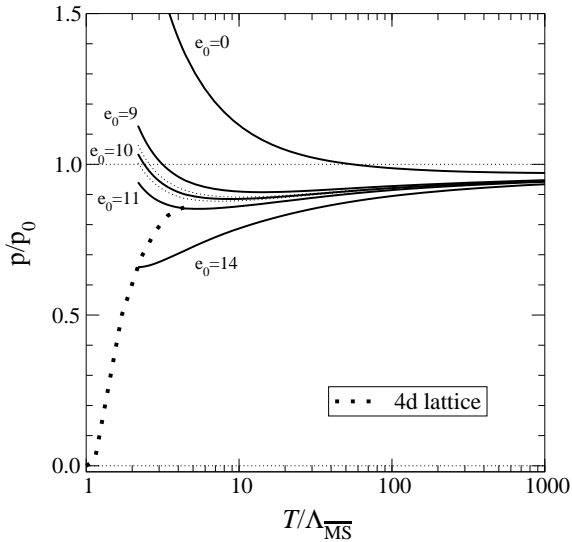}
\caption{The pressure of hot gluons,
calculated by dimensional reduction,
for various values of the undetermined constant $e_0$  \cite{Kajantie:2000iz}.
Statistical errors are shown for $e=10$.
The dotted line indicates predictions of lattice calculations.
}
\label{fig:dimred}
\end{figure}

In Fig.~\ref{fig:dimred} we see the range of predictions for the pressure of 
hot QCD with no quarks for various values of the constant $e_0$.
The value $e_0=10(1)$ 
agrees best with 4D lattice calculations. The effort to calculate
$e_0$ independently is currently underway.

This method is certainly better behaved than 4D perturbation theory.
It may also avoid some of the difficulties that beset
4D lattice calculations. For example, light quarks only come in through
the coefficients in the 3D action, and so do not impose
any additional computational burden. Also
it is possible to perform calculations at $\mu>0$
\cite{HLP}. But
it also inherits some of their limitations, such as being restricted
to thermodynamic observables that can be measured in Euclidean time.

\section{Lattice QCD at high temperature and low density}
\label{sec:hiden-latt}

Lattice calculations at non-zero density (chemical potential $\mu>0$)
have to contend with the ``sign problem''. Until recently this meant
that there were no lattice results for realistic
(3 color, 3+1 dimensional) QCD 
away from the $\mu=0$ axis of the phase diagram. 
There has been a lot of work on more tractable variants such as
QCD with two colors \cite{twocolor}, 
QCD at finite isospin density \cite{Kogut:2002zg}, 
Gross-Neveu models in 2+1 dimensions \cite{2+1GrossNeveu}
and so on. It has also been suggested that at asymptotically high
density the sign problem in QCD might become more tractable 
\cite{Hong:2002nn}.

In the last few years this situation 
has changed, with three groups reporting lattice results for
the position of the crossover line and even the critical point in the
high-$T$ low-$\mu$ region of the QCD phase diagram.
Before discussing those results, I will review the nature of the
sign problem.

The sign problem is not a deep problem of principle, 
but it is a very inconvenient
feature of the standard formulation of the QCD functional integral.
The problem is that when $\mu\neq 0$, the
factor of $\det(M[A])$ in the gauge field functional integral
\eqn{lattZ} is no longer a positive real number. This means that
the only practical way of evaluating such a high-dimensional integral,
the Monte-Carlo importance sampling method, which interprets the 
Boltzmann weight $B[A] = \det(M[A])\exp(-S_{\rm gauge})$ 
as a probability, cannot be applied. 

In principle it might seem that the sign problem could be 
straightforwardly fixed by ``reweighting''
\cite{Glasgow}. One example of this would be to
split off the phase of the weight by writing
$B[A] = \bigl|B[A]\bigr|\exp(i\phi[A])$, and
use the absolute value of the Boltzmann factor for importance 
sampling,
\beq
\ba{rcl}
Z_{||} &=& \int {\cal D}\!A \,\bigl|B[A]\bigr| \\[1ex]
\< O \>_{||} &=& (1/Z_{||}) \int {\cal D}\!A 
 \,O[A] \,\bigl|B[A]\bigr|
\ea
\eeq
The true expectation value of the observable 
can be obtained by including the complex phase in the observable,
taking its expectation value in the $|B|$ ensemble, and 
dividing by a reweighting factor
\beq
\< O \> = \< O \exp(i\phi[A]) \>_{||} / (Z/Z_{||}) .
\label{observable}
\eeq
But the reweighting factor is the ratio of two partition functions,
so it is related to the difference between their free energy densities $f$
\beq
Z/Z_{||} = \exp\bigl(-V(f-f_{||})\bigr) .
\eeq
The true partition function $Z$ contains the phase factor $\exp(i\phi[A])$,
and is therefore smaller than $Z_{||}$, so $f > f_{||}$.
Thus the reweighting factor in \eqn{observable} becomes exponentially
small in the thermodynamic limit $V\to\infty$.  The same is true for
$\< O \>_{||}$, because $\langle O \rangle$ itself is not
exponentially large in $V$.  This means exponentially large statistics
are required to obtain any given accuracy in $\<O\>$.

It must be stressed that the sign problem arises from the way we
structure our
evaluation of the functional integral, integrating out the fermion
fields for each fixed gauge field configuration. The physically relevant
quantities such as $Z$ and $\<O\>$ are real and positive,
and if we could restructure the ensemble of
gauge field configurations into families within which the weights
and observables analytically added up to positive values
then there would be no sign problem. This is what cluster algorithms
can do, and such a restructuring has been achieved for
the $\mathbb{Z}_3$ model that approximates QCD at very high quark 
mass and chemical potential \cite{Alford:2001ug}.
One strategy for the ultimate solution of the sign problem
is to recast the QCD functional integral into a form to 
which cluster algorithms can be applied  \cite{clusterQCD}.

In the meantime, it makes sense to try to perform calculations
using the methods available, in regions of the phase diagram
where the sign problem is weakest. This is the region of
small $\mu$ and high temperature.
In the last few years, various groups have 
have been exploring techniques applicable at $\mu \lesssim T$:
multi-parameter reweighting, derivatives with respect to $\mu$ at $\mu=0$,
and imaginary chemical potential.

\begin{figure}[htb]
\includegraphics[width=\hsize]{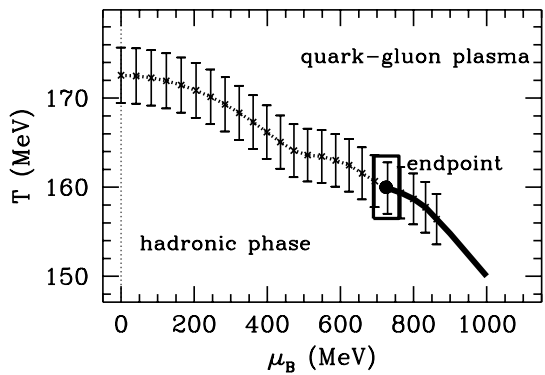}
\caption{The position in the $\mu$-$T$ plane of the crossover between
the hadronic and QGP phases, including an estimate of the position
of the critical point at $(\mu^*,T^*)$ and the first-order line,
obtained from a lattice
calculation with two-parameter reweighting \cite{FK-01}. Note that
the horizontal axis is $\mu_B = 3 \mu$.
}
\label{fig:mpw}
\end{figure}

\subsection{Multi-parameter reweighting.}
Fodor and Katz \cite{Fodor:2001au} suggested a reweighting algorithm like
the one described above, but using the $\mu=0$ ensemble to measure
observables rather than the $|B|$ ensemble.
They  exploit the freedom to vary the temperature
of the $\mu=0$ ensemble, and thereby improve its overlap
with the desired $\mu>0$ ensemble. This only postpones the sign problem
to larger volumes, but from their initial results it appears
that reasonable results may now be obtained on lattices that are
large enough to be physically interesting.
Their estimate of the crossover line and chiral critical point 
is shown in Fig.~\ref{fig:mpw} (for the equation of state
see \cite{Fodor:2002km}).
The numerical values should be interpreted with caution:
the light quark masses are well above their physical values,
so the true critical point is presumably at a somewhat lower
chemical potential \cite{Halasz:1998qr}.
The lattice quark action is unimproved (Kogut-Susskind) and
used at a relatively coarse lattice spacing $a\sim 0.2~\fm$
where it is known to have large discretization errors.
Thus there are potentially large systematic errors, and as
with all reweighting methods the statistical errors can be
misleadingly small.
Nonetheless, this is the only method that has obtained the
position of the chiral critical point, and it seems there is a real chance
that it may be able to do so for physical quark masses. 
This is a
striking improvement over the situation two years ago.

\subsection{{\boldmath $\p/\p\mu$} at {\boldmath $\mu=0$}.}
\begin{figure}[htb]
\includegraphics[width=\hsize]{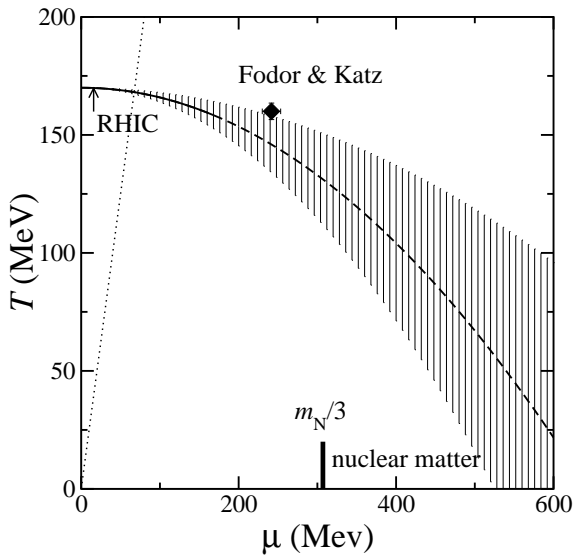}
\caption{
QCD hadronic/QGP crossover line from derivatives of susceptibilities
with respect to $\mu$, evaluated at $\mu=0$ \cite{Allton:2002zi}.
Note the excellent agreement with Fodor and Katz's multiparameter reweighting
calculation of the critical point $(\mu^*,T^*)$ (diamond).
}
\label{fig:dbydmu}
\end{figure}

A closely related strategy for extending lattice QCD
to small chemical potential is to
evaluate derivatives of observables
with respect to $\mu$ at $\mu=0$, where there is no sign problem
\cite{taro}. This is like taking the $\mu$-derivative of
Fodor and Katz's reweighting factor, and including that in the observable.
The relevant calculations have been performed by 
the Bielefeld/Swansea group \cite{Allton:2002zi}.
Their observables are the susceptibilities  for the Polyakov loop and chiral
condensate, the maxima of which indicate the position of the crossover
line in the $\mu$-$T$ plane. Using improved gauge and staggered quark actions
on a lattice of spacing $a\approx 0.2~\fm$, they obtain the results shown in
Fig.~\ref{fig:dbydmu}. The results agree very well with the
multi-parameter reweighting lattice calculation, whose estimate of
the critical endpoint is also shown.
It remains to be seen whether this approach has advantages over
multi-parameter reweighting. In particular, it is very hard to know
how deep into the $\mu>0$ region the Taylor expansion around $\mu=0$
will continue to work. To the left of the
dotted $\mu/T=0.4$ line in Fig.~\ref{fig:dbydmu} the sign problem
is under control, but the importance
of higher order $\mu$-derivatives is yet to be checked.

\subsection{Imaginary {\boldmath $\mu$}}
\begin{figure}[htb]
\includegraphics[width=\hsize]{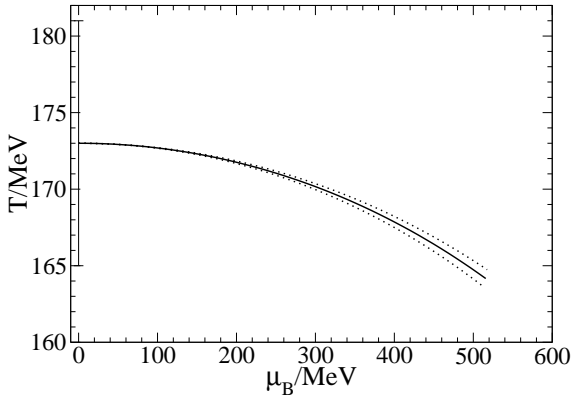}
\caption{QCD hadronic/QGP crossover line from analytic continuation
of imaginary $\mu$ results to real $\mu$ \cite{deForcrand:2002ci}.
The result is consistent with Figs.~\ref{fig:dbydmu},\ref{fig:mpw}.
Note that the horizontal axis is $\mu_B = 3 \mu$.
}
\label{fig:imagmu}
\end{figure}

Over the years, various workers have pointed out that the
sign problem only occurs for real chemical potential---the functional integral
with imaginary $\mu$ is amenable to standard Monte-Carlo methods
\cite{imagmu}. Now full 4D calculations by de~Forcrand and Philipsen
\cite{deForcrand:2002ci} have yielded estimates of the position
of the critical line  in the $\mu$-$T$ plane for two-flavor QCD
(for early explorations of 4 flavor QCD, see \cite{D'Elia:2002pj}).
These authors perform calculations at several imaginary values of $\mu$,
and then analytically continue to real $\mu$ by fitting their results
to a power series in $\mu^2$, and extrapolating from negative to positive
$\mu^2$ \cite{Lombardo:1999cz}.
The observables used to identify the crossover line were, as usual,
the susceptibilities for the plaquette, Polyakov loop, and chiral condensate.
Above the critical point ($\mu>\mu^*$) there is a line 
in the $\mu$-$T$ plane along which
these diverge in the large volume limit: this is the first-order line.
Below the critical point ($\mu<\mu^*$) each susceptibility
shows a ``ridge'' of finite height in the large volume limit, and
these ridges are expected to lie close together, approximately defining
a narrow crossover strip.
At large but finite volume,
the line defined by the ridge/divergence of each susceptibility $\chi$
will be an analytic function $T_\chi(\mu)$.
At infinite volume different susceptibilities
will give the same line above $\mu^*$ but may give different ones below 
$\mu^*$ so it is not clear that the function remains analytic for all $\mu$,
but at the volumes
studied so far $T_\chi(\mu)$ turns out to be a very docile expansion
in $\mu^2$ for imaginary $\mu$, fitting
very well to the form $c_0 + c_1 \mu^2$
with no need for a $\mu^4$ term. 
The resultant crossover line for QCD with two light flavors  is
plotted in Fig.~\ref{fig:imagmu}. It is consistent with the
results from the other methods, and the errors are smaller.

Imaginary chemical potential is quite different from the two methods
described above, and the errors appear to be well under control.
It will be very interesting to see whether it can estimate the
location of the critical point as accurately as it delineates the
crossover region.

\subsection{Relevance to experiment}

It is very impressive that the different methods described above,
all of which are being applied to  QCD for the first time,
achieve agreement in the position of the crossover line, and that there is
even a rough estimate of the position of the critical point
$\mu_B^*\sim 700~\MeV, T^*\sim 160~\MeV$. 
Putting aside caution
and taking the numbers seriously for a moment, could such a
critical point be accessible to heavy ion collision experiments?
Treating the estimated value of $\mu_B^*$ as an upper limit,
and bearing in mind that a heavy ion collision 
scans a range of temperatures down to the chemical freeze-out
temperature $T_{\rm fo}$
as the fireball expands, we can see 
from Table \ref{tab:hic} that there is a chance that
quark matter near the critical point could have been made
at SPS. It may also be produced at
the future experiment planned at GSI \cite{GSI-future}. And now we can 
even envisage the possibility that theorists may predict
the position of the critical point
before experimentalists observe it.

\begin{table}[htb]
\begin{tabular}{ccc}
\hline
experiment & $\mu_B$ & $T_{\rm fo}$ \\
\hline
RHIC & $40~\MeV$ & $170~\MeV$ \\
SPS  & $250~\MeV$ & $170~\MeV$ \\
AGS  & $550~\MeV$ & $130~\MeV$ \\
SIS  & $700~\MeV$ & $80~\MeV$ \\
\hline
\end{tabular}
\caption{ Approximate chemical freeze-out
temperatures and baryon chemical potentials
for heavy-ion collisions at various colliders \cite{BM-00}.}
\label{tab:hic}
\end{table}

\section{Far from equilibrium}

One motivation for calculations of the QCD equation of 
state and transport properties at
high temperature is their application to heavy ion
collisions at current and future heavy-ion colliders.
The success of hydrodynamic models of RHIC collisions
\cite{Heinz:2002un} indicates that
equilibrium is achieved within a very short time $\sim 1~\fm/c$,
but perturbative estimates give longer timescales \cite{MuellerSon}.
It is therefore very important to understand 
nonperturbative field theory
far from equilibrium.

\begin{figure}[htb]
 \includegraphics[width=\hsize]{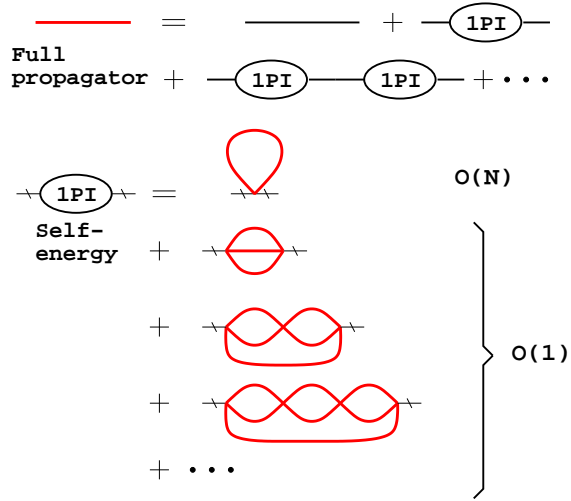}
\caption{Coupled NLO Schwinger-Dyson equations for the
full propagator from the 2PI effective action approach of
\cite{Berges:2000ur}.
}
\label{fig:2PI}
\end{figure}

In the last two years there has been significant progress
in treating non-equilibrium scalar field theory.
It is based on systematic loop \cite{Berges:2000ur} 
and $1/N$ \cite{Berges:2001fi,Aarts:2002dj} 
expansions of the two-particle-irreducible
(2PI) effective action formalism \cite{CJT-74,Phi}
discussed in section~\ref{sec:pert}.
The formalism is illustrated diagramatically in Fig.~\ref{fig:2PI}.
A typical mean-field calculation would only include the $\O(N)$ tadpole
(momentum-independent) contribution to the self-energy, and would not
show thermal equilibration at long times.
The NLO 2PI calculation also includes the $O(1)$ bubble chain contribution
of Fig.~\ref{fig:2PI},
which is crucial for including rescattering effects that lead to
equilibration.

\begin{figure}[htb]
 \includegraphics[height=\hsize,angle=+90]{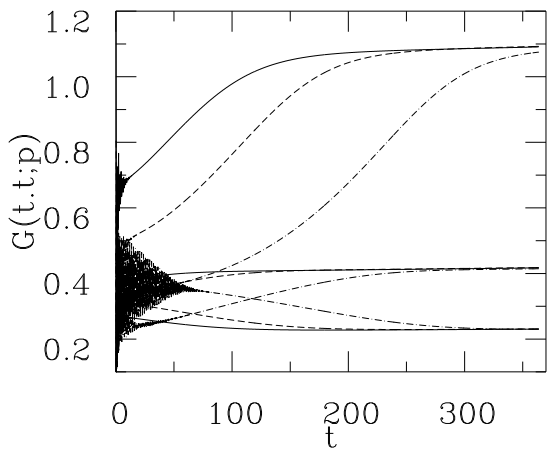}
\caption{Time dependence of three
modes of the equal time propagator in 
1+1D $\phi^4$ theory at non-zero temperature, for various
initial conditions. Calculated by 3-loop 2PI effective action
\cite{Berges:2001fi}. Note the quickly damped initial oscillations,
followed by drifting 
and then comparably slow approach to equilibrium.
}
\label{fig:nonequil}
\end{figure}

Fig.~\ref{fig:nonequil} shows the time evolution of the propagator $G(p)$
in 1+1D scalar field theory, for three momentum modes $|p|=0,3,5$. The
solid lines are for initial conditions close to the equilibrium value, the
dashed lines for initial conditions further from the equilibrium value, and
the dot-dashed lines are for initial conditions far from the equilibrium
value.

We see that the modes evolve in three stages,
\ben
\setlength{\itemsep}{-0.5\parsep}
\item Early: Initial oscillations quickly damped with rate
$\ga^{\rm(damp)}$.
\item Intermediate: Drifting towards the equilibrium value with 
non-exponential/power-law behavior.
\item Late: Exponential approach to equilibrium, with rate
$\ga^{\rm(therm)}\neq \ga^{\rm(damp)}$.
\een

The rate of approach to equilibrium $\ga^{\rm(therm)}$ is found to be
quite different from the initial damping rate $\ga^{\rm(damp)}$,
and the overall equilibration time in scalar theories is clearly much longer
than the microscopic timescales.
Obviously it will be very interesting to push these calculations to
higher dimensions \cite{Berges:2002cz}
where spontaneous symmetry breaking can be studied,
and to extend these methods to gauge theories, 
to see whether QCD predicts equilibration times as short as
$1~\fm/c$ for the QGP.

\section{Conclusions}

There is great enthusiasm and activity among theorists studying 
high-temperature and high-density QCD, stimulated by the successes of
current heavy-ion experiments and the anticipation of future ones, and
also by ongoing advances in the quality of observational data for
neutron stars.

In high-density QCD, the upsurge of interest in color
superconductivity is maturing into a effort to develop signatures for
the presence of color-superconducting quark matter in compact stars.
While ultra-high-density first-principles calculations and medium-density 
NJL calculations are becoming more sophisticated, useful results
also follow from parameterizing the quark pairing with a gap
parameter and studying its effects on phenomenologically
important quantities.

In high-temperature QCD, lattice gauge theory is making continuous
advances, achieving reliable calculations of equilibrium properties
of the QGP such as screening lengths and the equation of state
at temperatures up to a few times $T_c$.  Dimensionally reduced 
lattice calculations will be able to extend this to arbitrarily high
temperatures, when one remaining constant is computed.
Various resummed perturbative calculations appear to agree
with the lattice on the approximate size of 
deviations from the free equation
of state above the critical temperature, although
they are not yet on a solid independent footing.
Complete leading-order
perturbative expansions of transport
properties, including all leading logs,
are being achieved. The photon emission rate of
the QGP has been computed, and other quantities will
soon be calculated.

Lattice QCD has at last taken its first step into the non-zero-density
region of the phase diagram. Remarkably, there is already agreement
among the different methods on the position of the crossover line,
and there is even a first estimate of the position of the chiral
critical point, which is a sharply-defined physical observable.
It would therefore be valuable for theorists, as well as
striving for more accurate predictions of the position of the
critical point, to further develop the existing work
on signatures of the critical point in heavy-ion collisions.

Finally, field theories far from equilibrium are becoming accessible
to controlled calculation. The 2PI effective action method has
applications to disoriented chiral condensates, critical
fluctuations, and other complexities of heavy ion collisions, as well
as cosmological questions such as the exit from inflation in the early
universe. Efforts to apply this method to higher-dimensional scalar field
theories, spontaneous symmetry breaking, and gauge theories are in progress.

Strong interaction physics at high temperature and density is a
promised land that lies at the heart of modern physics, bordering on
particle physics, nuclear physics, astrophysics, and condensed matter
physics.  Our first explorations are revealing the true extent of its
fertility and richness.


\end{document}